\documentclass[preprint,aps,prd,nofootinbib,superscriptaddress]{revtex4}
\usepackage{hyperref}
\usepackage[pdftex]{graphicx,color}
\usepackage{slashed}
\usepackage{amsmath,amssymb}
\usepackage{hhline}
\usepackage{subfigure}
\usepackage{epstopdf}

\newcommand{\beq}{\begin{equation}}
\newcommand{\eeq}{\end{equation}}
\newcommand{\beqa}{\begin{eqnarray}}
\newcommand{\eeqa}{\end{eqnarray}}

\newcommand{\Bbar}{\,\overline{\!B}{}}
\newcommand{\Dbar}{\,\overline{\!D}{}}
\newcommand{\Kbar}{\,\overline{\!K}{}}
\def\B0bar{\Bbar{}^0}
\def\D0bar{\Dbar{}^0}
\def\K0bar{\Kbar{}^0}

\newcommand{\br}{{\cal B}_{D^*\to e^+e^-}}

\begin{document}
\begin{flushright}
SI-HEP-2015-18\\
QFET-2015-24\\
WSU-HEP-1503
\end{flushright}
\title{\boldmath Direct probes of flavor-changing neutral currents \\
in $e^+e^-$-collisions}

\author{Alexander Khodjamirian}
\author{Thomas Mannel}
\affiliation{Theoretische Physik 1, Naturwissenschaftlich-Technische 
Fakult\"at,\\
Universit\"at Siegen, D-57068 Siegen, Germany}

\author{Alexey A.\ Petrov}
\affiliation{Theoretische Physik 1, Naturwissenschaftlich-Technische 
Fakult\"at,\\
Universit\"at Siegen, D-57068 Siegen, Germany}

\affiliation{Department of Physics and Astronomy\\
        Wayne State University, Detroit, MI 48201, USA}

\affiliation{Michigan Center for Theoretical Physics\\
        University of Michigan, Ann Arbor, MI 48196, USA}


\begin{abstract}
\noindent
We propose a novel method to study flavor-changing neutral currents 
in the $e^+e^-\to D^{*0}$  and $e^+e^-\to B_{s}^* $ transitions, tuning the energy of 
$e^+e^-$- collisions to the mass of the narrow vector 
resonance $D^{*0}$ or $ B_{s}^*$. We present a thorough 
study of both short-distance and long-distance contributions 
to $e^+e^-\to D^{*0}$ in the Standard Model and investigate 
possible contributions of new physics in the charm sector. 
This process, albeit very rare, has clear advantages  
with respect to the 
$D^0 \to e^+e^-$ decay: the helicity suppression is absent, 
and a richer set of effective operators can be probed.  
Implications of 
the same proposal for $B_{s}^*$ are also discussed.
\end{abstract}

\maketitle

\section{Introduction}\label{Intro}

Experimental studies of the flavor-changing neutral currents (FCNC) are among the most promising ways to reveal 
virtual effects of possible new physics (NP) in heavy meson decays. The FCNC transitions have been thoroughly studied in the 
$b$-quark sector, where the Standard Model (SM) is seen to dominate the decay amplitudes \cite{CMS:2014xfa,Bobeth:2013uxa}.
Nonetheless, recent hints at anomalies in the exclusive $B\to K^{(*)}\ell^+\ell^-$ decays \cite{LHCb:BKll,LHCb:Rk} call for additional studies in the 
$b$-flavor FCNCs, preferably, with observables having  less hadronic uncertainties than in the rare semileptonic decays.
The leptonic decays $B_{s,d}\to \ell^+\ell^-$ remain the ``cleanest'' probes, however, only for the axial-vector/pseudoscalar effective 
operators. Moreover, due to helicity suppression, only the $\ell=\mu,\tau$ modes are accessible in leptonic decays, leaving the 
detection of the electron modes extremely challenging.

Still, even if the $b\to s(d)\ell^+\ell^-$ transitions are conform with the SM, the situation might be different in the 
charm-quark FCNCs, where ample room for NP effects in the  $c\to u \ell^+\ell^-$ transitions  is  available 
(see, e.g., \cite{Artuso:2008vf}). Rare decays of the type $D^0 \to \ell^+\ell^-$ for $\ell = \mu, e$ 
have a potential to probe a variety of NP scenarios. 
The studies of these decays are seemingly appealing because 
for the local operators mediating these transitions, both in 
SM and beyond, all nonperturbative effects are 
accumulated in a single parameter, the $D^0$ decay constant. 
However, in SM the long-distance effects dominate the rare decays of charmed mesons. 
It is extremely difficult to estimate these contributions model-independently. Moreover, since the initial state in the  $D^0 \to e^+e^-$
decay is a pseudoscalar meson, the helicity suppression again
makes the observation of this process very difficult.
This feature persists in many NP models as well. 
In addition,  the rare radiative decay $D^0 \to \gamma e^+e^-$ has 
a branching ratio that is enhanced by a factor of ${\cal O}(\alpha m_D^2/m_e^2)$ 
compared to $D^0 \to e^+e^-$, since the additional photon lifts the helicity 
suppression \cite{Fajfer:2002gp}. If the emitted photon is soft, this
transition can be easily misidentified as $D^0 \to e^+e^-$, which 
further complicates its experimental observation. 

An interesting alternative to $D^0 \to e^+e^-$ process that is {\it not} 
helicity-suppressed is a related decay $D^{*}(2007)^0 \to e^+e^-$.
While also probing the FCNC $c\bar{u} \to \ell^+\ell^-$ transition, this decay 
is sensitive to the contributions of operators that 
$D^0 \to \ell^+\ell^-$ cannot be sensitive to. Unfortunately, 
a direct study of the $D^* \to e^+e^-$  decay is practically impossible, 
since the $D^*$  decays strongly or electromagnetically.

\begin{figure}[t]
\begin{center}
\includegraphics[width=8cm]{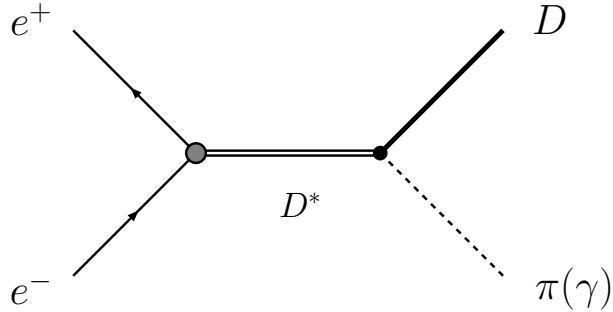}
\end{center}
\caption{\label{figure} \it Probing the $ c\bar{u}\to e^+ e^-$ vertex
with the $D^*(2007)^0$ resonance production in $e^+e^-$ collisions .}
\end{figure}

Nevertheless, as we shall argue in this paper, it might be possible to probe
the $D^*\to e^+e^-$ transition experimentally. Assuming time-reversal 
invariance, it would be equivalent to measure the corresponding production process 
$e^+e^- \to D^*$, as shown in Fig.~\ref{figure}. In order to do so, we 
propose to run an $e^+e^-$ collider, such as BEPCII \cite{Zweber:2009qf}, at the 
center-of-mass energy corresponding to the mass of the $D^*$ meson. Note that 
BEPCII already scanned this region of energies, achieving the luminosity
of about  $5 \times 10^{31}$ cm$^{-2}$ s$^{-1}$ around $\sqrt{s}=2$ GeV \cite{RoyPC}.
If produced, the $D^{*0}$ resonance will decay via strong 
($D^{*0}\to D^0\pi^0$) or electromagnetic ($D^{*0}\to D^0\gamma$)  interactions  
with branching fractions of $(61.9\pm 2.9)\%$ and 
$(38.1\pm 2.9)\%$ respectively.\footnote{Note that the charged mode 
$D^{*0} \to D^+\pi^-$ is forbidden by the lack of the available phase space.}

In the setup discussed in this paper, the $D^*$ production process is very rare. However, the 
identification of even a single charmed-meson final state from the 
$e^+e^-\to D^{*0} \to D^0\pi^0$ decay would provide an unambiguous tag for this flavor-changing 
production process. Naturally, one also needs an adequate quality of the $\pi^0\to 2 \gamma$ 
identification and pion-kaon separation in the $D^0$ decays in order to reject background 
processes. 

Our proposal may also be realized in the $b$-quark sector by scanning the region of the 
$B^*_{s,d}$ resonances at an $e^+ e^-$ collider. This will probe the processes 
$ e^+e^-\to B_{s,d}^*$ originating from the $b\to s(d)\ell^+\ell^-$ quark currents. 
In fact, studying the transitions involving electrons could also shed some light on recent 
hints at lepton non-universality in $b \to s e^+e^-$ versus  $b \to s \mu^+ \mu^-$ \cite{LHCb:Rk}.   

Note also that tuning an $e^+e^-$ accelerator to the
masses of resonances is not the only possibility to access their 
production. Some sensitivity to these processes could be also 
achieved by studying radiative return events at currently running 
$e^+e^-$ machines operating at their nominal energies.

The rest of this paper is devoted to a more detailed discussion of this proposal
and to the relevant theoretical estimates.

\section{$e^+e^-\to D^*$ resonant production.}\label{Prod}

Let us consider a generic scattering amplitude of $e^+e^-\to D \pi$, and assess the contribution of the
narrow resonance $D^*$ to this process depicted in Fig.~\ref{figure}. 
Writing this amplitude as a matrix element of a generic lepton-quark interaction
\beq\label{eq:H}
{\cal H} = \frac{\lambda^\prime}{M^2} (\bar c \gamma_\mu u)(\bar{e}\gamma^\mu e), 
\eeq
where only the vector currents are kept for simplicity and 
an effective scale $M$ and dimensionless coupling $\lambda^\prime$
are introduced, we obtain 
\beqa
{\cal M}(e^+e^-\to D^0 \pi^0) &=& \langle D^0(p_D)\pi^0(p_\pi) | {\cal H} |e^+(p_+)
e^-(p_-)\rangle
\nonumber\\
&=& \frac{\lambda^\prime}{M^2}\langle D^0(p_D)\pi^0(p_\pi)|\bar c \gamma_\mu u |0\rangle 
\langle 0 |\bar{e}\gamma^\mu e|e^+(p_+)e^-(p_-) \rangle
\\
&=&
\frac{\lambda^\prime}{M^2} \big(2 f^+_{D^0\pi^0}(s) p_{\pi_\mu}\big) \
\overline v (p_+)\gamma ^\mu u(p_-)\,,
\label{eq:Meff}
\nonumber
\eeqa
where the lepton current is factorized out and the hadronic matrix element 
is expressed via the $D^0\to\pi^0$ vector form factor at $s =(p_D+p_\pi)^2\geq (m_D+m_\pi)^2$. Note that
$\sqrt{s}$  is the center-of-mass energy of the $e^+e^-$-collision.  
Up to an isospin factor, the same vector form factor appears in the semileptonic 
$D^0\to \pi^-\ell^+\nu_\ell$ decay, where $s \leq (m_D-m_\pi)^2$. 

At first place, it is the effective coupling in Eq.~(\ref{eq:H}) that 
determines the value of the cross section calculated from (\ref{eq:Meff}). Yet, 
the presence of a narrow  resonance in the form factor is also crucial. 
To see that, we isolate the two lowest resonance contributions to the form factor, 
that is, $D^*$ and $D^{*'}=D(2600)$, where the latter, with the mass $m_{D^{*'}}=2612\pm 6$ MeV 
and total width $\Gamma_{D^{*'}}=93\pm 14$ MeV \cite{PDG},
is the most suitable candidate for the first radial excitation of $D^*$-meson. 
In the resulting decomposition
\beq
f^+_{D^0\pi^0}(s)= 
\frac{f_{D^{*0}} g_{D^{*0}D^0\pi^0} m_{D^{*0}}}{2(m_{D^{*0}}^2-s-im_{D^{*0}}\Gamma_0)}+
\frac{f_{D^{*0\prime}} g_{D^{*0\prime}D^0\pi^0}m_{D^{*0\prime}}}{2(m_{D^{*0\prime}}^2-s-im_{D^{*0\prime}}\Gamma_{D^{*0\prime}})}+
\left[f^+_{D^0\pi^0}(s)\right]_{\rm bgr}\,,
\label{eq:formfDpi}
\eeq
we expressed the residues of both poles via decay constants of vector resonances and 
their strong couplings  to $D\pi$, defined, respectively, as 
\begin{eqnarray}
\langle 0| \overline u \gamma^\mu c | D^*(p) \rangle = f_{D^*} m_{D^*} \epsilon^\mu (p)\,,
\label{DstarDC}\\
\langle D^0(p_D)\pi^0(p_\pi) 
|D^{*0}(p)\rangle = -g_{D^{*0}D^0\pi^0}(\epsilon(p)
\cdot p_\pi)\,,
\end{eqnarray}
where $\epsilon^\mu(p)$ is the $D^*$ polarization vector, $p=p_D+p_\pi$, and 
we isolated the background contribution $\left[f^+_{D^0\pi^0}(s)\right]_{\rm bgr}$ with respect to the two resonances.

While experimentally only an upper bound 
for the total width of $D^{0*}$, $\Gamma_0 < 2.1$~MeV,  
is available \cite{PDG}, we can compute the actual value of this width
from the measured total width of the charged $D^{+*}$ meson,
$\Gamma_+=83.4\pm 1.8$ keV \cite{PDG}. Using the isospin 
symmetry to relate the strong $D^*D\pi$ couplings and 
taking into account the phase space correction (see, e.g.,\cite{Belyaevetal}) 
we obtain:
\beqa
\Gamma_0&=& 
\Gamma (D^{* 0} \to D^0 \pi^0)
+ \Gamma (D^{* 0} \to D^0 \gamma)
\nonumber
\\
&\simeq&
\frac{\Gamma_+{\cal B}_{D^{*+} \to D^0\pi^+}}{2 }
\left (\frac{\lambda(m^2_{D^{*0}},m^2_{D^0},m^2_{\pi^0})}{\lambda(m^2_{D^{*^+}},m^2_{D^0},m^2_{\pi^+})}  \right)^{\!\!3/2}
\!\!\bigg(1+ \frac{{\cal B}_{D^{*0}\to D^0\gamma} }{{\cal B}_{D^{*0}\to D^0\pi^0}}\bigg)
\simeq
60 ~\mbox{keV}\,,
\label{eq:Dstarwidth}
\eeqa
where ${\cal B}_{D^*\to f} = \Gamma(D^*\to f)/\Gamma_0$ is the branching fraction 
of $D^* \to f$,  $\lambda(x,y,z)$ is the kinematic K\"allen function and we employed 
the data on branching fractions from \cite{PDG}, neglecting small 
experimental errors.

The Breit-Wigner form used in Eq.~(\ref{eq:formfDpi}) is certainly applicable for such a narrow resonance as $D^*$. We adopt the same approximation for the broad resonance $D^{*'}$, which is sufficient for an order-of-magnitude estimate. At $s=m_{D^*}^2$ the magnitude of the excited state contribution to 
the form factor is suppressed with  respect to the $D^{*0}$-pole term by the factor, approximately,
\beq
\Bigg|\frac{f_{D^{0*\prime}}\,g_{D^{*'0}D^0\pi^0} m_{D^{*0\prime}}}{f_{D^{0*}}\,g_{D^{*0}D^0\pi^0} m_{D^{*0}}}\Bigg|
\times \Bigg|\frac{i\Gamma_0}{2\Delta-i\Gamma_{D^{*'}}}\Bigg|
\sim 5.0\cdot 10^{-5}\,,
\label{eq:ratio}
\eeq   
where $\Delta= m_{D^{*0'}}-m_{D^*}\simeq 600 $ MeV and the $O(\Delta/m_{D^*})$ terms 
are neglected. In the above, we assume that 
the strong couplings and 
decay constants of the radially excited and ground $D^*$ states are in the same ballpark, so that the first factor in this ratio is of $O(1)$. 
In fact, the ratio of decay constants in Eq.~(\ref{eq:ratio}) is less than one, as 
the QCD sum rule  predictions indicate \cite{Gelhausen:2014} .
Our estimate also implies that if one produces a resonance with a width of $O(100-150)$ MeV,
typical for the light vector mesons such as $\rho$ or $K^*$, there is no relative gain in the 
resonance cross section. We emphasize that the very small width of $D^{*}$, driving up 
the cross section, essentially originates from the miniscule phase space of 
its single strong decay mode. In this situation, the
radiative mode of $D^*$ becomes equally important and the FCNC transition acquires a tiny but 
non-negligible branching fraction.   

Having in mind a strong suppression of all other than $D^*$ 
contributions to Eq.~(\ref{eq:Meff}) at $\sqrt{s}\simeq m_{D^*}$, 
the cross section of $e^+e^- \to D\pi$ 
can be written in a standard resonance form:
\begin{eqnarray}\label{BW}
\sigma(e^+e^- \to D\pi)_{\sqrt{s}\simeq m_{D^*}} \equiv \sigma_{D^*}(s) 
= \frac{12 \pi}{m_{D^*}^2} 
\ {\cal B}_{D^* \to e^+e^-} {\cal B}_{D^*\to D\pi} \
\frac{m_{D^*}^2 \Gamma_0^2}
{(s-m_{D^*}^2)^2+m_{D^*}^2\Gamma_0^2},
\end{eqnarray}

To exploit the $D^*$-resonance enhancement around  $\sqrt{s} = m_{D^*}$, described
by Eq.~(\ref{BW}), we tacitly assume that an appropriate tuning of the electron-positron
accelerator beams can be performed, so that their energy resolution is 
smaller than the spread of the resonance  cross section.
Then we can simply use Eq.~(\ref{BW}) at $s = m_{D^*}^2$, yielding
\begin{equation}
\sigma_{D^*}(m^2_{D^*}) 
=  {\cal B}_{D^* \to e^+e^-}  
{\cal B}_{D^*\to D\pi}
\frac{12 \pi}{m_{D^*}^2} \ 
  \simeq {\cal B}_{D^* \to e^+e^-}(2.26\times 10^6)\,\mbox{nb}\,.  
\end{equation}
Let us recall that the total cross section $\sigma(e^+e^-\to hadrons)$ is about 50 nb 
at $\sqrt{s}=2.0 $ GeV \cite{PDG}. 

The expected number of $e^+e^-\to D\pi$ events at $\sqrt{s}=m_{D^*}$ 
is given by the product
\beq 
N_{D^*}=\sigma_{D^*}(m^2_{D^*})\, \epsilon {\small \int} Ldt   
\label{eq:Nofev}
\eeq
where $\epsilon$  and $\int L dt$ are the detection efficiency and 
time-integrated luminosity, respectively.
The condition $N_{D^*}\geq 1$ leads to a lower bound on the $D^* \to e^+e^-$ branching
fraction  that  still allows one to detect the process $e^+e^-\to D^*\to D\pi$: 
\begin{equation}
\label{Constr}
{\cal B}_{D^* \to e^+e^-}\geq \left(\frac{1}{\epsilon \int Ldt } \right)
\times \frac{m_{D^*}^2}{12 \pi \ {\cal B}_{D^*\to D\pi}}.
\end{equation}
For example, an average $e^+e^-$ luminosity at the level 
of $L \approx 1.0\times 10^{32} ~{\mbox cm}^{-2}s^{-1}$,
with a "Snowmass year" ($\sim 10^7 s$) of running at the $D^*$ resonance
yields $\int L dt = 1.0$ fb$^{-1}$. 

Under these conditions, the single-event sensitivity implied by the 
bound (\ref{Constr}) 
means that branching fractions 
\beq
\label{eq:boundBR}
{\cal B }_{D^* \to e^+e^-}> 4\times 10^{-13}  
\eeq
could in principle be probed. 
While the above bound is a very crude estimate of the possible sensitivity, 
as we did not take into account detection efficiency, assuming $\epsilon=1$,
it can serve as a useful criterion. In order to see if and when this bound can be approached,
let us first calculate the branching fraction  $\br $ in the SM. We 
need this quantity as it is the key parameter determining 
the $e^+e^-\to D^*$ cross section.

\section{$D^* \to e^+e^-$  transition }\label{Eval}

The most general expression for the $D^*\to e^+e^-$ decay amplitude 
can be written as
\begin{widetext}
\begin{eqnarray}\label{DstarAmp}
A(D^*\to e^+e^-) = \overline{u}(p_-, s_-) \left[
A \gamma_\mu + B \gamma_\mu \gamma_5 
+ \frac{C}{m_{D^*}} (p_+-p_-)_\mu 
\right. ~~~~~~~~~~~~~~
\nonumber \\
\qquad + \left.
\frac{D}{m_{D^*} }(p_+-p_-)_\mu i\gamma_5 \
\right] v(p_+,s_+) \ \epsilon^\mu(p),
\end{eqnarray}
\end{widetext}
where $A$, $B$, $C$, and $D$ 
are dimensionless constants which absorb the underlying effective quark-lepton 
interaction and the vacuum $\to D^*$ hadronic matrix 
elements (e.g., the $D^*$ decay constant). Note that the simplified effective interaction
(\ref{eq:Meff}) used in the previous section corresponds to 
$A=(\lambda^\prime/M^2)f_{D^*}m_{D^*}$ and all other constants put to zero.
The amplitude (\ref{DstarAmp}) leads to the branching fraction
\begin{eqnarray}\label{BRDstar}
\br
= \frac{m_{D^*}}{12 \pi\Gamma_0}
\left[
\left(\left|A\right|^2 + \left|B\right|^2\right) +
 \frac{1}{2} \left(\left|C\right|^2 + \left|D\right|^2\right)
\right],
\end{eqnarray}
where we neglected the mass of the electron. 

It is straightforward to compute the SD
part of the $D^*\to e^+e^-$ amplitude triggered by a generic 
effective Hamiltonian, not necessary containing only the SM operators. 
The decay amplitude is
\begin{equation}\label{Amplitude}
\langle e^+e^- | {\cal H}_{\rm eff} | D^* \rangle = G \sum_i c_i (\mu) 
\langle e^+e^- | \widetilde Q_i | D^* \rangle|_\mu\,,
\end{equation}
where $G$ is a constant with the dimension of inverse squared mass 
that sets the scale for the operators.
For example, $G=4G_F/\sqrt{2}$ in the SM or $G=1/\Lambda^2$ for new physics.
In the above, $\widetilde Q_i$ are the effective operators of dimension six, and 
$c_i$ are the corresponding (dimensionless) Wilson coefficients.
The most general set of ten local operators producing the $c \bar{u} \to\ell^+ \ell^-$ 
transitions reads \cite{Golowich:2009ii}
\begin{eqnarray} \label{SetOfOperatorsLL}
\widetilde Q_1 &=& (\overline{\ell}_L \gamma_\mu \ell_L) \ 
(\overline{u}_L \gamma^\mu c_L), 
\ \
\widetilde Q_4 = (\overline{\ell}_R \ell_L) \ 
(\overline{u}_R c_L),
\nonumber \\ 
\widetilde Q_2 &=& (\overline{\ell}_L \gamma_\mu \ell_L) \ 
(\overline{u}_R \gamma^\mu c_R),
\ \
\widetilde Q_5 = (\overline{\ell}_R \sigma_{\mu\nu} \ell_L) \ 
( \overline{u}_R \sigma^{\mu\nu} c_L),
\nonumber \\ 
\widetilde Q_3 &=& (\overline{\ell}_L \ell_R) \ (\overline{u}_R c_L),  
\end{eqnarray}
with five additional operators $\widetilde Q_6, \dots, \widetilde Q_{10}$ 
obtained respectively from those in Eq.~(\ref{SetOfOperatorsLL}) by 
the substitutions $L \to R$ and $R \to L$. We will be using the set of operators 
$\widetilde Q_{i=1,...,10}$ to assess possible 
contributions of any generic new physics model
to $D^* \to e^+e^-$.

The operators with (pseudo)scalar quark currents 
$ \widetilde Q_{3-4}$ and $ \widetilde Q_{8-9}$
do not contribute to this process. On the other  hand,
contrary to the $D^0\to \ell^+\ell^-$ decay, not only the vector operators
 $ \widetilde Q_{1-2}$ and $ \widetilde Q_{6-7}$ but also the tensor operators
$\widetilde Q_5$ and $\widetilde Q_{10}$ could, in principle, contribute to $D^*\to e^+e^-$. 
However, in most of NP models, these effective operators 
are absent, motivating us to neglect their contributions. 
This is equivalent to 
setting $C=D=0$ in Eq.~(\ref{DstarAmp}). In terms of the Wilson
coefficients $c_i$ the constants $A$ and $B$ are
\begin{eqnarray}\label{AB}
A &=&~\frac{G}{4} f_{D^*} m_{D^*}  \left(c_1+c_2+c_6+c_7\right),
\nonumber \\
B &=&-\frac{G}{4} f_{D^*}m_{D^*}  \left(c_1+c_2-c_6-c_7\right).
\end{eqnarray}
The effective operators in SM (see \cite{Burdman:2001tf} for the definition) 
mediating the $D^*\to e^+e^-$  transition in SM are easily matched to the 
operator set in Eq.~(\ref{SetOfOperatorsLL}),
\begin{equation}\label{O9O10}
O_9 = \frac{e^2}{16\pi^2}\left(\widetilde Q_1 + \widetilde Q_7\right)\,,
~~
O_{10} = \frac{e^2}{16\pi^2}\left(\widetilde Q_7 - \widetilde Q_1\right)\,.
\end{equation}
The expressions for the corresponding Wilson coefficients 
$C_9^{\rm c}$ and $C_{10}^c$,
where a superscript $c$ indicates that CKM matrix elements are 
included in their definitions, can also be found in \cite{Burdman:2001tf}. 

In addition to $O_9$ and $O_{10}$, in SM the $D^* \to e^+e^-$ 
amplitude receives a contribution from the magnetic dipole operator $O_7$ 
coupled to the leptons via virtual photon. The corresponding part of the 
effective Hamiltonian reads
\begin{equation}
H^{(7\gamma)}_{\rm eff}= \frac{4 G_F}{\sqrt{2}} C_7^{\rm c, eff}
\left(\frac{e}{16\pi^2} m_c\ \overline u_L \sigma^{\mu\nu} c_R F_{\mu\nu}\right)\,.
\end{equation}
A calculation of this 
contribution requires the knowledge of  
the tensor (transverse) decay constant of the $D^*$:
\begin{equation}\label{DstarDCT}
\langle 0| \overline u \sigma^{\mu\nu} c | D^*(p) \rangle = i f^T_{D^*} 
\left( \epsilon^\mu p^\nu-p^\mu \epsilon^\nu\right)\,.
\end{equation}
In the absence of the estimate of $f^T_{D^*}$, we rely 
on the properties of light vector mesons for which the vector
(longitudinal) and  tensor decay constants 
are in the same ballpark and simply assume  that 
$f^T_{D^*}=f_{D^*}$. One can now calculate 
the SD contribution to $D^* \to e^+e^-$ in the SM
in terms of effective constants:
\begin{eqnarray}\label{ABsm}
A^{(SD)} &=&\frac{\alpha}{2\pi}\frac{G_F}{\sqrt{2}} f_{D^*} m_{D^*}  
\left[C_9^{\rm c, eff} + 2 \frac{m_c}{m_{D^*}} \frac{f^T_{D^*}}{f_{D^*}} C_7^{\rm c, eff}\right],
\nonumber \\
B^{(SD)} &=&\frac{\alpha}{2\pi} \frac{G_F}{\sqrt{2}} f_{D^*} m_{D^*}  C_{10}^c\,,
\end{eqnarray}
which results in the SD contribution to the branching fraction:
\begin{equation}\label{SDbrSM}
 \br= \frac{\alpha^2G_F^2}{96 \pi^3\Gamma_0}
m_{D^*}^3 f_{D^*}^2 \left(\left|C_9^{\rm c, eff}
+ 2 \frac{m_c}{m_{D^*}} \frac{f^T_{D^*}}{f_{D^*}} C_7^{\rm c, eff}
\right|^2+ \left|C_{10}^c\right|^2\right).
\end{equation}
Adopting $m_c=1.3$ GeV, we use for the Wilson 
coefficient\footnote{We thank Dirk Seidel 
for providing us this coefficient 
calculated at NLL order.}, $C_9^{\rm c}(\mu=m_c)=0.198|V_{ub}^*V_{cb}|$,  
neglect $C_{10}^{\rm c}(\mu=m_c)$
and  employ the results of the two-loop calculation \cite{Greub:1996wn}
for the remaining effective coefficient,
$C_7^{\rm c, eff}(\mu=m_c)=-0.0025$. We also use
the central value of the QCD sum rule estimate 
$f_{D^*} \approx 242$ MeV \cite{Gelhausen:2013wia}.
Substituting all input paramaters in Eq.~(\ref{SDbrSM}), we find
\beq
\br^{SD} \approx 2.0 \times 10^{-19}\,.
\label{eq:sdsm}
\eeq
As expected, this number is extremely small, several orders of magnitude
below the lowest accessible branching fraction Eq.~(\ref{eq:boundBR})
and thus beyond any realistic experimental setup.
Still, it is instructive to remind the reader that 
the short-distance width of the similar decay of the pseudoscalar $D^0$
is many orders of magnitude smaller: ${\cal B}^{SD}_{D^0 \to e^+e^+} \sim 10^{-23}$,
whereas  ${\cal B}^{SD}_{D^0 \to  \mu^+\mu^-} \sim 10^{-18}$ (see, e.g., \cite{Paul:2010pq}).   

\section{ Long-distance contributions}

Generally, in rare charm decays, a significant enhancement of the 
decay rate is expected in SM due to LD
contributions, 
generated by the four-quark weak interaction
combined with the emission of the $e^+e^-$-pair via virtual photon. 
It is very difficult to reliably estimate these contributions in  
$D\to \ell^+\ell^- $ decay because the  two-photon intermediate state 
overlaps with long-distance hadronic interactions. 

To investigate the case of $D^* \to e^+e^-$ decay,
we isolate the relevant $\Delta C=1$, single Cabibbo-suppressed 
transitions in the effective Hamiltonian  ${\cal H}_w$ of the SM, representing
it in a form of the two four-quark operators:
\begin{eqnarray}
\label{eq:Hw}
{\cal H}_w = \frac{4 G_F}{\sqrt{2}} \sum_{q=d,s} \Bigg[ 
\left(C_1^{c(q)}+\frac{C_2^{c(q)}}{N_c}\right)\ (\overline q_L \gamma_\nu q_L)  (\overline u_L \gamma_\nu c_L)
\nonumber \\ 
+  2  C_2^{c(q)} (\overline q_L \gamma_\nu T^a q_L)  
(\overline u_L \gamma_\nu T^a c_L)
\Bigg]\,,
\end{eqnarray}
where $T^a$ are the color-octet matrices.
Note that we again included relevant CKM matrix elements into the 
definition of $C_1^{c(q)}$ and  $C_2^{c(q)}$.

\begin{figure}[t]
\begin{center}
\includegraphics[width=7cm]{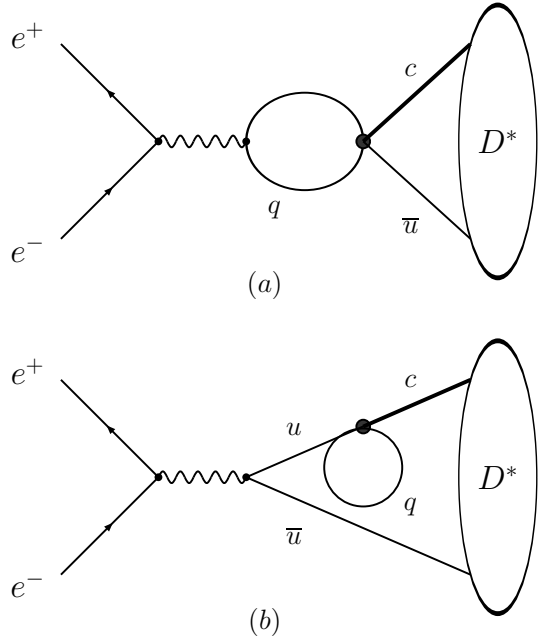}
\end{center}
\caption{\label{figure2} \it 
 Long-distance contributions  to $e^+ e^-\to D^{*0}$ caused by 
the virtual photon interaction (a) with $q=d,s$  quark in factorizable approximation,
forming the polarization operator;  (b) with $u$ quark.
}
\end{figure}

The LD contribution to the $D^* \to e^+e^-$  
decay amplitude is given by the matrix element 
\begin{equation}\label{eq:LDampl}
\langle e^+e^- | {\cal H}_{w} | D^*(p) \rangle=
-e^2\bar{u}(p_-,s_-)\gamma^\mu v(p_+,s_+)
\left(\frac{\Sigma_\mu(p^2)}{p^2}\right)\Bigg|_{p^2=m_{D^*}^2}\,,
\end{equation}
where we factorized out the lepton current, the photon propagator
and defined the hadronic matrix element in a form of the 
correlation function
\begin{equation}\label{Correlator}
\Sigma_\mu (p^2) = i \int d^4 x e^{ip\cdot x}
\langle 0 | T \left\{ j^{em}_\mu(x) {\cal H}_w(0) \right\} | D^*(p) \rangle\,,
\end{equation}
where $j^{em}_\mu=\sum_{q=u,d,s,c} Q_q\overline q \gamma_\mu q $ is the 
electromagnetic (e.m.) quark current. The quark-level 
diagrams  corresponding to this amplitude are depicted 
in Fig.~\ref{figure2}, where we distinguish two main topologies
corresponding to 
the virtual photon interacting with (a) the $q=d,s$  quarks (upper panel)
and (b) with the $u$ quark (lower panel).   
Both contributions develop imaginary parts
due to the intermediate light-hadron states.
We neglect the e.m. interaction with the heavy $c$-quark.

To estimate the LD contribution corresponding to the 
diagram Fig.~\ref{figure2}a we adopt  factorization approximation, 
that is, neglect the gluon exchanges 
between  $\bar{q}q$ and $\bar{u}c$ fields in ${\cal H}_{w} $.
Hence, we retain in Eq.(\ref{Correlator}) 
only the product of color-neutral vector currents 
and obtain:
\begin{widetext}
\begin{eqnarray}\label{Correlator1}
\Sigma^{(a)}_\mu (p^2) 
&=&  \frac{G_F}{\sqrt{2}}  \sum_{q=d,s} Q_q\left(C_1^{c(q)}+\frac{C_2^{c(q)}}{N_c}\right)
\left\{
i \int d^4 x e^{i p \cdot x}
\langle 0 | T \left\{ \overline q \gamma_\mu q (x) \
\overline q \gamma_\nu q (0) \right\} | 0 \rangle 
\right\}
\nonumber\\
&\times&\langle 0 |  \overline u \gamma^\nu c | D^*(p) \rangle\,, 
\end{eqnarray}
\end{widetext}
where we recognize the quantity in the curly braces as the 
polarization tensor $\Pi_{\mu\nu}^{(q)}$ for the $q$-flavored vector current.
Substituting its decomposition 
\begin{equation}
\Pi_{\mu\nu}^{(q)} (p) = \left(-g_{\mu\nu} p^2 + p_\mu p_\nu \right) \Pi^{(q)}(p^2)
\end{equation}
in Eq.~(\ref{Correlator1}) and parametrizing the matrix element of 
the $\bar{u}c$ current according to Eq.~(\ref{DstarDC}), we finally obtain
the LD transition amplitude in the form (\ref{DstarAmp}) with 
\begin{eqnarray}\label{LDamp}
A^{(LD,a)} = 4\pi\alpha Q_q \frac{G_F}{\sqrt{2}} f_{D^*} m_{D^*} 
\left(C_1^{c}+\frac{C_2^{c}}{N_c}\right)
\Big( \Pi^{(d)}(p^2) -\Pi^{(s)}(p^2)\Big)\Bigg|_{p^2=m_{D^*}^2},
\end{eqnarray}
and $B^{(LD,a)}=0$. In the above expression, we slightly modified the Wilson 
coefficients, so that 
$C_{1,2}^{c}\equiv C_{1,2}^{c(s)}\simeq -C_{1,2}^{c(d)} $, taking into account that the
CKM factors 
for the $s$- and $d$-quark parts of ${\cal H}_{w}$ are approximately equal 
and have opposite signs. 
As expected, in the flavor $SU(3)$ limit  the whole 
LD contribution vanishes, reflecting the GIM cancellation.

 Note that the factorizable quark-gluon effects are implicitly retained 
in the two separate hadronic quantities, the decay constant and 
the polarization operator. Nonfactorizable QCD corrections will involve 
contributions of the color-octet and axial-vector quark 
currents. We expect that these corrections are 
suppressed either by $\alpha_s(m_c)$ or by the powers
of $\Lambda_{QCD}/m_c$, since the characteristic 
momenta flowing through the quark loop in 
the diagram of Fig.\ref{figure2}a are of order of $m_D\sim m_c$. 
For the same reason, one may also argue that 
the mixing between $\bar{s}s$ and $\bar d d$ loops  
in the polarization operator is suppressed, which is also 
in line with the  OZI suppression valid in the vector-meson channel. 
To substantiate the  expected suppression of higher-order effects, 
one has to calculate the nonfactorizable multi-loop diagrams explicitly,
a technically difficult task we postpone to future studies. 
  
Returning to the LD contribution (\ref{LDamp}) we  
estimate first the polarization operators.
Note that $\Pi^{(q)}(p^2)$ satisfies a (once-subtracted) dispersion relation
\begin{equation}
\Pi^{(q)}(p^2) = \frac{p^2}{12\pi^2Q_q^2} \int_0^\infty ds \frac{R^{(q)}(s)}{s(s-p^2-i\epsilon)}\,,
\label{eq:Rqq}
\end{equation}
where $R^{(q)}(s)$ is the normalized $e^+e^-$ cross section to hadrons 
initiated by the quark flavor $q=s,d$, so that below the charm-anticharm 
threshold
\begin{equation}
\label{eq:R}
R(s) \equiv \frac{\sigma(e^+e^-\to hadrons)}{\sigma(e^+e^-\to \mu^+\mu^-)}=\sum_{q=u,d,s} R^{(q)}(s)\,.
\end{equation}
In principle, $R_q(s)$ could be extracted from the experimental data. 
In practice, however, it is hard to disentangle hadronic states generated by 
the $d$ and $s$-quark currents. Therefore, for the sake of estimate, we 
employ  the parametrization of $R_q(s)$ stemming from the QCD sum rule
analysis \cite{SVZ} of the $u,d$- and $s$-quark correlation functions
and based on the quark-hadron duality:
\begin{eqnarray}
R^{(d)}(s) &=& 12 \pi^2 Q_d^2\Big[\sum\limits_{V=\rho^0,\omega}\frac{f_V^2}{2}\delta(s-m_V^2) 
+ \frac1{4\pi^2}\left(1+\frac{\alpha_s}{\pi}\right)\theta(s-s_0^{d})\Big]\,,
\label{eq:PolQQd}
\\
R^{(s)}(s) &=& 12 \pi^2 Q_s^2\Big[f_\phi^2\delta(s-m_\phi^2) 
+ \frac1{4\pi^2}\left(1+\frac{\alpha_s}{\pi}\right)\theta(s-s_0^{s})\Big] \,,
\label{eq:PolQQs}
\end{eqnarray}
where the decay constants of the light vector mesons are defined
via matrix elements of the electromagnetic current:
\beq
\langle 0|j_\mu^{em}|V\rangle = \kappa_{V}m_Vf_V\epsilon_{(V)}^{\mu} \,.
\eeq
Here the coefficients $\kappa_{\rho}=1/\sqrt{2}$, $\kappa_{\omega}=1/(3\sqrt{2})$ and 
 $\kappa_{\phi}=-1/3$ reflect the valence quark content of these mesons.
Furthermore, we take into account the total widths of the 
vector mesons, replacing delta-functions in Eqs.~(\ref{eq:PolQQd}-\ref{eq:PolQQs}) 
with the Breit-Wigner approximations.

To calculate the difference of the polarization operators 
entering  Eq.~(\ref{LDamp}), we use $f_\rho= 220 $ MeV, $f_\omega=197 $ MeV, and $f_\phi=228 $ MeV, obtained from the 
experimental values of the $\rho,\omega$ and $\phi$ leptonic widths \cite{PDG}, 
We also choose the effective thresholds \cite{SVZ}:
$s^{d}_0 = 1.5 $ GeV$^2$ and  $s^s_0 = 1.95 $ GeV$^2$.
To finalize our estimate of the LD amplitude,
we take the relevant combinations of 
CKM parameters $\eta_{CKM} \equiv V^*_{us}V_{cs}\simeq -V^*_{ud}V_{cd}=\lambda(1-\lambda^2/2)$,
where $\lambda=0.22537$ \cite{PDG} is the Wolfenstein parameter, 
and use the Wilson coefficients at $\mu=m_c=1.3$ GeV at LO(NLO):  
 $C_{1}^{c}/\eta_{CKM}=-0.53(-0.41)$ and $C_{2}^{c}/\eta_{CKM}= 
1.28(1.21)$.

If we assume that the LD contribution of the annihilation type calculated above 
saturates alone the decay amplitude, the resulting branching fraction of the $D^*\to e^+e^-$ decay, turns out quite sensitive to the mutual cancellation of the 
Wilson coefficients  $C_{1}$ and $C_2/3$: 
\beq
\br^{LD,A} \simeq
\left\{ 
\begin{array}{l}
4.7\times 10^{-20} ~~\mbox{(NLO)} \\
5.7 \times 10^{-18} ~~\mbox{(LO)}
\end{array} \right. \,.
\label{eq:BLD}
\eeq
The cancellation is numerically 
less pronounced in LO, in which case the above branching fraction grows 
by almost two orders of magnitude with respect to the NLO case, 
becoming substantially larger than the SD width (\ref{eq:sdsm}).  

One could also try to estimate the difference of polarization
operators in Eq.~(\ref{LDamp}) by subtracting the $d$- and $s$-quark loop diagrams
from each other. This is equivalent to replacing $R^{(q)}(s)$ 
by the parton spectral density (the second term 
in brackets in Eqs.~(\ref{eq:PolQQd}-\ref{eq:PolQQs})) and integrating from 
$4m_q^2$ to infinity. In this case, to account for a correct normalization
of the effective operators in $H_{\rm eff}$ and to avoid  infrared unstable 
terms of  $O(\ln(m_d/m_s))$  in the difference of the polarization
operators, one has to add a (scheme dependent) constant term to the dispersion 
integral, which contains $\ln m_q^2/\mu^2$. The resulting difference
of the loop functions is then proportional to $(m_s^2-m_d^2)/\mu^2$ as it should
be in the GIM cancellation.\footnote{In the hadronic representation of the 
polarization operator used above, the same cancellation can be traced 
in the difference of QCD sum rules for $f^2_{\rho,\omega}$ and $f^2_{\phi}$ and stems 
from the difference of loop diagrams in the perturbative part and, in addition, from 
the $O({m_s\langle\overline{s}s\rangle-m_d\langle\overline{d}d}\rangle)$  
terms originating from the vacuum condensate contributions.} 
The simple loop estimate yields, instead of Eq.~(\ref{eq:BLD}),
\beq\label{eq:BLDloop}
\br^{LD,A} \simeq 
\left\{ 
\begin{array}{l}
2.0\times 10^{-20} ~~\mbox{(NLO)} \\
2.4 \times 10^{-18} ~~\mbox{(LO)}
\end{array} \right. \,,
\eeq
in the same ballpark as the estimate (\ref{eq:BLD}).

Turning to the LD contribution  
represented by the diagram in Fig.~\ref{figure2}\,b, 
we present a rough estimate 
of its amplitude, taking the imaginary part and
accounting only 
for the lowest $\pi^+\pi^-$ and $K^+K^-$ 
intermediate states. 
Other contributions, including multiparticle ones, are possible. Those can in principle give 
even larger amplitudes due to phase space suppression of the $SU(3)$-related 
intermediate states \cite{Falk:2001hx}.

According to the unitarity condition, the imaginary part 
of the $D^*\to e^+e^-$ amplitude generated by these particular 
intermediate states is expressed via the $e^+e^-\to P^+P^-$ e.m. 
amplitude which contains the  
$\langle 0 | Q_u\overline{u}\gamma_\mu u |P^+P^-\rangle $ 
form factor taken at the timelike momentum transfer $p^2=m_{D^*}^2$.
The e.m. amplitude is multiplied by the  
$D^*\to P^+P^-$ nonleptonic amplitude ($P=\pi,K$) and integrated over
the $P^+P^-$ phase space. 

Applying the naive factorization approximation for the 
nonleptonic amplitude, we introduce the relevant hadronic matrix element 
\beq
p^+_\alpha\langle P^-(p_-)|\overline{q}\gamma^\alpha\gamma_5 c|D^*(p)\rangle   
= 2i(\epsilon(p)\cdot p_+)A_0^{D^*\to P}(p^2=m_{D^*}^2)m_{D^*}\,,
\label{eq:A0formf}
\eeq
which depends on the one particular form factor of the $D^*\to P$ transition, 
(we define these form factors analogous to the well familiar 
$D\to V$ form factors).
Using also the decay constant 
of the light pseudoscalar meson:
\beq 
\langle P^+(p_+)|\overline{u}\gamma^\rho\gamma_5 q|0\rangle = -ip^\rho_+f_P\,, 
\label{eq:fP}
\eeq
we obtain the LD contribution in the following form:
\begin{eqnarray}
\mbox{Im}A^{(LD,b)}= \frac{\alpha}{12} \frac{G_F}{\sqrt{2}}
\left(C_2^c+\frac{C_1^c}{N_c}\right)f_\pi A_0^{D^*\pi}(m_\pi^2) 
Q_uF_\pi^{em}(m_{D^*}^2)\beta_\pi^3m_{D^*} 
\nonumber\\
\times
\left(1-\frac{f_K \beta_K^3}{f_\pi \beta_\pi^3}\frac{A_0^{D^*K}(m_K^2) 
F_K^{em}(m_{D^*}^2)}{A_0^{D^*\pi}(m_\pi^2)F_\pi^{em}(m_{D^*}^2)}\right)\,,
\label{eq:LDpeng}
\end{eqnarray}
where $\beta_P=\sqrt{1-4m_P^2/m_{D^*}^2}$. For an accurate numerical 
estimate we need to calculate the form factors of $D^*\to P$ transition,
which is beyond our scope, demanding, e.g., a dedicated application of 
QCD light-cone sum rules. 
For an order-of-magnitude estimate 
we assume that these form factors are in the same ballpark 
as the (correspondingly normalized) $D\to \pi,K$  form factors and 
take $A_0^{D^*\pi}(m_\pi^2)\sim 1$. The pion e.m. form factor 
$F_\pi^{em}(m_{D^*}^2)\simeq 0.28$ is known from a 
measurement \cite{BaBarFpi} in the time-like momentum region. 
For the SU(3)-violating ratio of the e.m. 
and heavy-light form factors  we conservatively assume that they can 
vary within $\pm 30 \%$
\beq
\frac{A_0^{D^*K}(m_K^2) F_K^{em}(m_{D^*}^2)}{A_0^{D^*\pi}(m_\pi^2)F_\pi^{em}(m_{D^*}^2)}= 1\pm 0.3\,.  
\label{su3ratio}
\eeq
After that we obtain the lower limit on the branching fraction
assuming that the LD contribution (\ref{eq:LDpeng}) is dominant:
\beq
\br^{(LD,b)} \geq (0.1- 5.0)\times10^{-19}\,,
\label{eq:BLDpeng}
\eeq
where the broad interval mainly reflects the variation within 
the limits assumed in \ref{su3ratio}, whereas 
switching from NLO to LO Wilson coefficients
is a minor effect in this case.    

A more complete analysis of LD effects including possible interference
between contributions of the two types considered above as well  
as a more complete set of intermediate states is an interesting 
task we postpone for the future. The estimates presented above 
demonstrate possible approaches which could be developed
further. 

The branching fractions presented in Eqs.(\ref{eq:BLD}),
(\ref{eq:BLDloop}) and (\ref{eq:BLDpeng}),
can achieve the level of $10^{-18}$, that is, not very much larger than 
the typical values ~(\ref{eq:sdsm}) in the presence of only SD effects. 
Altogether, the probability of $e^+e^-\to D^{*0}$ in SM remains 
considerably smaller than the estimated
minimal level (\ref{eq:boundBR}) reachable (optimistically) by experiment. 
Therefore,  if observed or at least constrained at the 
level of Eq.~(\ref{eq:boundBR}), the $e^+e^-\to D^*$  events  
will definitely have no SM background. 
While this is basically true for a majority of FCNC transitions
in the charm sector, such as $D\to \ell^+\ell^-$ or semileptonic decays, 
the $e^+e^-\to D^*$ transition has a clear advantage 
of having a relatively moderate LD background.

\section{$D^*\to e^+e^-$ transitions and new physics.}\label{NP}

It is interesting to estimate what possible NP scale could be probed by the processes 
discussed in this paper. As follows from Eq.~(\ref{SetOfOperatorsLL}), there are ten possible 
operators that parameterize any NP contribution to any $c\bar{u} \to e^+e^-$ process. 
Some of those operators can be probed both in $D^0 \to e^+e^-$ and $D^{0*} \to e^+e^-$
transitions. Some can only be reached in $D^{0*} \to e^+e^-$ or (more challenging from 
hadronic point of view) rare semileptonic charm decays $D \to M e^+e^-$. 
It would be interesting to compare the possible reach of $D^0$ and $D^{0*}$ decays.

Assuming that a NP contribution is dominated by a single operator, it is 
easy to see from Eqs.~(\ref{BRDstar}) and (\ref{AB}) that 
\begin{equation}
\Lambda \sim \left(
\frac{1}{3\pi} \frac{m_{D^*}^3 f_{D^*}^2}{32 \Gamma_0} \frac{C^2}{\br}
\right)^{1/4},
\end{equation}
where $C=\left|c_i\right|$ for $i=1,2,6$, or $7$. Employing the
upper bound (\ref{eq:boundBR}) we find that 
observation of a single event in a ``Snowmass year" of running would probe NP scales of 
the order of $\Lambda \sim 2.7$ TeV provided that $C \sim 1$.
Let us compare this to the NP reach of $D^0 \to e^+e^-$ decay for the same operators.
Using \cite{Golowich:2009ii}, 
\begin{equation}
\Lambda \sim \left(
\frac{m_{D} m_e^2 f_D^2}{32 \pi \Gamma_D} \frac{C^2}{{\cal B}_{D\to e^+e^-}}
\right)^{1/4},
\end{equation}
where $\Gamma_D$ is the total width of the $D^0$ meson,
and the current experimental bound, ${\cal B}_{D\to e^+e^-} = 7.9\times 10^{-8}$, 
we find that only scales $\Lambda \sim 200$ GeV are currently probed by $D\to e^+e^-$ decay.
It is the presence of the lepton mass factor that severely limits the 
NP scale sensitivity in this process. 

To exemplify this discussion, let us consider two particular models of NP to see how well
they can be probed in $D^* \to e^+e^-$ transition. The selected models are 
by no means unique. The chosen examples simply illustrate the differences in 
sensitivities between $D^0 \to \ell^+ \ell^-$ and $D^* \to e^+e^-$.

{\em R-parity violating (RPV) SUSY.}
R-parity violating SUSY models can be probed in FCNC charm decays
\cite{Golowich:2009ii,Burdman:2001tf}. The relevant part of the 
superpotential can be written as 
\begin{eqnarray} \label{lrpar}
W_{\lambda'} &=& \tilde\lambda'_{ijk}
\left\{
V_{jl}\left[ 
\tilde\nu^i_L
\bar d^k_Rd^l_L+\tilde d^l_L\bar d^k_R\nu^i_L
+(\tilde d^k_R)^*(\bar\nu^i_L)^cd^l_L
\right]  
\right.
\nonumber \\
&-& \left.  \tilde e^i_L\bar d^k_Ru^j_L
-\tilde u^j_L\bar d^k_Ru_L^j-(\tilde d_R^k)^*(\bar e^i_L)^cu_L^j
\right\},
\end{eqnarray}
where the coupling parameters $\tilde\lambda'_{ijk}$ are 
defined such that $i$ denotes a generation number for 
leptons or sleptons, $j$ -- for up-type quarks, and  
$k$ -- for down-type quarks or squarks. As can be seen from 
Eq.~(\ref{lrpar}), the relevant FCNC quark transition 
$c + {\bar u} \to e^+ e^-$ is mediated by a tree-level $d$-squark 
exchange. Since its mass is much larger than the energy scale 
at which $D^*$ decay takes place, it can be integrated out, 
resulting in the effective Lagrangian 
\cite{Golowich:2009ii,Burdman:2001tf},
\begin{eqnarray}
& & 
{\cal L}_{\rm eff}^{\not R_p} = \frac{\tilde{\lambda}'_{12k}
\tilde{\lambda}'_{11k}} {2 m^2_{\tilde{d}^k_R}}\,
\widetilde Q_1 \ \ ,
\end{eqnarray}
which implies that $G c_1 = \tilde{\lambda}'_{12k}
\tilde{\lambda}'_{11k}/(2 m^2_{\tilde{d}^k_R})$ in Eq.~(\ref{Amplitude}).
Extracting the coefficients $A$ and $B$ and using Eq.~(\ref{AB}) 
leads to the branching fraction 
\begin{eqnarray}\label{RPV}
& &  
\br^{\not R_p}=\frac{1}{384 \pi} \frac{m_{D^*}^3 f_{D^*}^2}{m^4_{\tilde{d}^k_R} \Gamma_0}
\left| \tilde{\lambda}'_{12k} \tilde{\lambda}'_{11k} \right|^2 .
\end{eqnarray}
Contrary to the case of $D^0 \to \ell^+ \ell^-$, no helicity suppression
(factors of $m_\ell^2/m_D^2$) is seen in Eq.~(\ref{RPV}),
which in principle makes this process more sensitive to the NP parameters.
This feature exists for any NP model that is represented by 
$V\pm A$ interactions. Numerically, taking updated bounds on 
$\tilde{\lambda}'_{12k} \tilde{\lambda}'_{11k}$ from \cite{Petrov:2007gp},
conservatively (see also \cite{Fajfer:2007dy}),
\begin{equation}
\left|\tilde{\lambda}'_{12k} \tilde{\lambda}'_{11k}\right| \leq 3.83 \times 10^{-3}
\left[\frac{m_{\tilde{d}^k_R}}{300 \ \mbox{GeV}}\right],
\end{equation}
we estimate that $\br^{\not R_p}< 1.7 \times 10^{-14}$, implying that some 
prospects exist for improvement on this bound using the described process.

{\em Models with $Z^\prime$-mediated gauge interactions.}
Another interesting and representative model that we would like to 
consider here is a model with flavor-changing $Z^\prime$-mediated 
interactions. In general, 
\begin{eqnarray}\label{Zprime}
{\cal L}_{Z^\prime} &=& 
-g_{Z^\prime1}^\prime \overline \ell_L \gamma_\mu \ell_L Z^{\prime\mu} 
-g_{Z^\prime2}^\prime \overline \ell_R \gamma_\mu \ell_R Z^{\prime\mu} 
\nonumber \\
&-& g_{Z^\prime1}^{cu} \overline u_L \gamma_\mu c_L Z^{\prime\mu} 
-g_{Z^\prime2}^{cu} \overline u_L \gamma_\mu c_L Z^{\prime\mu} .
\end{eqnarray}
For $m_{Z^\prime}\gg m_D$ the Lagrangian in Eq.~(\ref{Zprime}) leads to
\begin{eqnarray}\label{ZprimeEff}
{\cal L}_{\rm eff}^{Z^\prime} &=& 
-\frac{1}{M_{Z^\prime}^2} \left[
g_{Z^\prime1}^\prime g_{Z^\prime1}^{cu} \widetilde Q_1 +
g_{Z^\prime1}^\prime g_{Z^\prime2}^{cu} \widetilde Q_2  + 
g_{Z^\prime2}^\prime g_{Z^\prime2}^{cu} \widetilde Q_6 +
g_{Z^\prime2}^\prime g_{Z^\prime1}^{cu} \widetilde Q_7 
\right].
\end{eqnarray}
Again, identifying the Wilson coefficients $c_i$  from Eq.~(\ref{ZprimeEff}) 
and computing $A$ and $B$ leads to the following branching fraction,
\begin{eqnarray}\label{BrZprime}
\br^{Z^\prime} &=& 
\frac{1}{12 \pi} \frac{m_{D^*}^3 f_{D^*}^2}{M_{Z^\prime}^4 \Gamma_0}
\left| g_{Z^\prime1}^{cu} + g_{Z^\prime2}^{cu}\right|^2
\left(
\left| g_{Z^\prime1}^\prime \right|^2 + \left| g_{Z^\prime2}^\prime \right|^2
\right) .
\end{eqnarray}
As with our previous example, Eq.~(\ref{BrZprime}) does not exhibit helicity 
suppression of the rate. Most importantly, $\br^{Z^\prime}$ is non-zero for purely 
vectorial interactions of the $Z^\prime$, which will be realized if, for example,
$ g_{Z^\prime1}^{cu} = g_{Z^\prime2}^{cu}$. This is contrary to 
$D^0 \to \ell^+\ell^-$ decay rate, where such vectorial couplings are
forbidden by vector current conservation \cite{Golowich:2009ii}.
There are five parameters that describe generic $Z^\prime$ interactions
with quarks and leptons, $g_{Z^\prime 1}^{cu}$, $g_{Z^\prime 2}^{cu}$,
$g^\prime_{Z^\prime 1}$, $g^\prime_{Z^\prime 2}$, and $M_{Z^\prime}$. To
assess the sensitivity of the $e^+e^-\to D^*$ production mechanism to $Z^\prime$
models numerically, let us make two simplifying assumptions. First, let us
assume that $Z^\prime$ only couples to left-handed
quarks,\footnote{Equivalently, we could have assumed that $Z^\prime$ only
couples to the right-handed currents. Then $g_{Z^\prime 1}^{cu}=0$ and
constraints would be obtained for $g_{Z^\prime 2}^{cu}$.} which would mean
that $g_{Z^\prime 2}^{cu}=0$. Second, let us assume that the $Z^\prime$ has
SM-like diagonal couplings to leptons,
\beq
g^\prime_{Z^\prime 1} =
\frac{g}{\cos\theta_W}\left(-\frac{1}{2}+\sin^2\theta_W\right), \qquad
g^\prime_{Z^\prime 2} = \frac{g \sin^2\theta_W}{\cos\theta_W},
\eeq
where $g$ is the SM $SU(2)$ gauge coupling. The branching fraction would
then only depend on the combination $g_{Z^\prime 1}^{cu}/M_{Z^\prime}^2$,
\beq
\br^{Z^\prime} = \frac{\sqrt{2} G_F}{3\pi \Gamma_0} m_{D^*}^3 f_{D^*}^2
\frac{\left|g_{Z^\prime 1}^{cu}\right|^2}{M_{Z^\prime}^2}
\frac{M_Z^2}{M_{Z^\prime}^2}
\left(\frac{1}{4}-\sin^2\theta_W + 2 \sin^4\theta_W\right).
\eeq
Taking the constraint $M_{Z^\prime}/\sqrt{g_{Z^\prime 1}^{cu}}>8.7\times 10^2$ GeV
from $D^0 \to \mu^+\mu^-$ \cite{Golowich:2009ii} yields
\beq
\br^{Z^\prime} < 2.5 \times 10^{-11},
\eeq
which is far above the SM predictions for this rate. 

\section{Implications for $B^*_{s}$ FCNC decays.}\label{Bstuff}

Similarly to $D^0 \to e^+e^-$, the $B_{s(d)} \to e^+e^-$ decay
is helicity-suppressed. Hence, it might be interesting to see if a $e^+e^-$
production process can be used to probe the $B^*_{s(d)} \to e^+e^-$ transitions.
Hereafter we will concentrate on the $B_s^*$ resonant production, 
the corresponding process with $B_d^*$ is very similar, but CKM suppressed  
and has therefore less chances to be detected. 
There are two important differences between the beauty and the charm case. 
First of all, the strong decays of the $B^*_s$ are kinematically forbidden, 
and the dominant channel  is the radiative one, 
$B_{s}^{0*} \to B^0_{s}\gamma$. This feature is welcome because it leads to  
a smaller  total width than for $D^*$. However, additional 
challenges might emerge for triggering the final state of the produced  $B_{(s)}^{*}$. Second, 
in the SM the  $B^{*}_{s} \to e^+e^-$ decay is dominated by 
the SD contributions, stemming from the well defined effective Hamiltonian. 
Hence one may expect a reasonable accuracy in predicting the 
decay rate, also because the only hadronic parameter involved is the $B^*_s$
decay constant.  On the other hand, one also has to assess the possible 
LD contributions, e.g., the effect similar 
to the one shown in Fig. \ref{figure2}a, but with the $c$-quark polarization operator. 

For $e^+e^-\to B_s^*$, we use the resonant cross section
\beq
\sigma_{B^*_s}(m^2_{B^*_s}) 
=  {\cal B}_{B_s^* \to e^+e^-}  
\frac{12 \pi}{m_{B^*_s}^2} \ 
\label{eq:crosssectB}
\eeq
and find that with the branching ratios  
\beq
{\cal B}(B^*_s\to e^+e^-)> 2.0\times 10^{-12}\,,
\label{eq:BRBgam}
\eeq 
the resonance production can be observed with at least one event.
Here we again assume one year running at the $B_s^*$ resonance energy,
with the luminosity of $\sim 10^{32}$ cm$^{-2}$ s$^{-1}$ and $\epsilon\sim 1$ 
detection efficiency. 
While no $e^+e^-$ collider is currently operating at this energy, future upgrades of 
the existing $e^+e^-$ machines, including also the radiative return setup, 
might make this energy region available for experimental studies.

Before calculating the branching fraction of $B^*_s\to e^+e^-$, 
it is important to fix the total width of $B^*_s$  which practically 
coincides with the  width of the single flavor-conserving 
radiative decay:
\beq 
\Gamma^{tot}_{B_s^{*}}\simeq  \Gamma(B_s^{*}\to B_s\gamma)=
\frac{\alpha}{24} |g_{B_s^{*}B_s\gamma}|^2
\left(\frac{m_{B_s^{*}}^2-m_{B_s}^2}{m_{B_s^{*}}}\right)^3\,.
\label{eq:Bgamma}
\eeq
Here we use the definition of the $B_s^*B_s\gamma$ coupling:
\beq
\langle B_s(p)\gamma(q) 
|B_s^{*}(p+q)\rangle =\sqrt{4\pi\alpha}\,g_{B_s^{*}B_s\gamma}\, \varepsilon^{\mu\nu\rho\lambda}
\epsilon^{*(\gamma)}_\mu q_\nu\epsilon^{(B_s^*)}_\rho p_\lambda\,,
\label{eq:gBBgam}
\eeq
The dominant contribution to  the $H^*\to H\gamma$ ($H=D,B$)  transitions 
stems from the long-distance photon emission off the light quark in 
the heavy meson. The analyses of these couplings in terms of 
heavy-hadron ChPT and related approaches \cite{DstDgamHQ,Colangelo:1993zq} allow
for a simplified parametrization of the coupling
\footnote{ The $H^*\to H\gamma$ couplings 
are also estimated\cite{DstDgamSR}  from QCD sum rules, 
 relating them to the 
electromagnetic susceptibility of QCD vacuum, the latter parameter is however
known with a rather large uncertainty.} :
\beq
g_{H^*H \gamma}\simeq \frac{Q_Q}{m_{H^*}}+ \frac{Q_q}{\mu_q}\,,
\eeq
where  $Q_{Q(q)}$  is the charge factor of 
the heavy(light) quark $Q=c,b$ ($q=u,d,s$) in $H^{(*)}$, and $\mu_q$ is a 
nonperturbative parameter, which does not scale with 
the heavy mass. Numerically, this relation describes well the 
two experimentally measured $D^{*0,+}\to D\gamma $ widths,
if $\mu_{u,d}\simeq 420-430$  MeV is taken. Using the same value
of $\mu_{u,d}$ for the $g_{B^{*0}B^0\gamma}$ coupling, 
we obtain $\Gamma(B^{*0}\to B^0\gamma)\sim 0.2$ keV, in the ballpark
of the estimates obtained in \cite{DstDgamHQ,Colangelo:1993zq}. 
To account for the $SU(3)$-flavor  symmetry violation, we 
adopt the model of \cite{Colangelo:1993zq} where the photon emission 
from the light-quark is described
via vector-meson dominance, so that $\mu_s\simeq \mu_{u,d}(m_\rho^2/m_\phi^2)$. 
We obtain then from Eq.~(\ref{eq:Bgamma})
\beq 
\Gamma^{tot}_{B_s^{*}}\simeq 0.07~ \mbox{keV}\,.   
\label{eq:Bstwidth}
\eeq
Our conclusion is that the $B_s^*$-resonance 
is considerably narrower than $D^*$. 

To estimate the probability of $B^*_s\to e^+e^-$ in SM, we employ   
the relevant SD part of the effective Hamiltonian for the $b\to s\ell^+\ell^-$
transitions:
\begin{eqnarray}
{\cal H}_{\rm eff} = -\frac{4 G_F}{\sqrt{2}}  V_{tb} V^*_{ts}\sum_{i=7,9,10}
C_i O_i + h.c.\,,
\end{eqnarray}
involving the operators: 
\begin{eqnarray}
O_7 &=& -\frac{e m_b}{16 \pi^2} \ \overline q_L \sigma^{\mu\nu} b_R F_{\mu\nu},
\nonumber \\
O_9 &=& \frac{e^2}{16 \pi^2} \left(\overline q_L \gamma^\mu b_L\right)
\left(\overline \ell \gamma_\mu \ell\right),
\\
O_{10} &=& \frac{e^2}{16 \pi^2} \left(\overline q_L \gamma^\mu b_L\right)
\left(\overline \ell \gamma_\mu\gamma_5 \ell \right).
\nonumber 
\end{eqnarray}
Using Eq.~(\ref{SDbrSM}) with the obvious substitutions of 
charm meson parameters by the beauty ones, we obtain for 
the branching fraction:
\begin{equation}\label{SDbrSM_forB}
{\cal B} (B_s^* \to e^+e^-) = \frac{\alpha^2G_F^2}{96 \pi^3\Gamma^{tot}_{B^*_s}}
m_{B_s^*}^3 f_{B_s^*}^2 \left| V_{tb} V^*_{ts}\right|^2
\left(\left|C_9
+2 \frac{m_b}{m_{B_s^*}} \frac{f^T_{B_s^*}}{f_{B_s^*}} C_7^{\rm eff}
\right|^2+ \left|C_{10}\right|^2\right).
\end{equation}

To refine our estimate, we also should add the nonlocal contribution
generated by the combination of the current-current weak operator
$O_{1,2}$ and the quark e.m. current. One of the 
important effects is the lepton pair emitted from the 
intermediate $\bar{c}c$ pair, described by  
the diagram similar to Fig.\ref{figure2}a. In 
the factorizable approximation,
this contribution is estimated in a full analogy with the LD
effect for the charmed vector-meson leptonic decay, i.e., we can use the same 
expression but replace the difference of $d$ and $s$ polarization operators   
with the single $c$-quark one: 
\begin{eqnarray}\label{LDBamp}
A^{(LD)} = 4\pi\alpha Q_c \frac{G_F}{\sqrt{2}}V_{cb}V^*_{cs}f_{B_s^*} m_{B_s^*} 
\left(C_1+\frac{C_2}{N_c}\right)\Bigg[-\frac{1}{4\pi^2}\left(\ln\frac{m_c^2}{\mu^2}+
1\right)+ \Pi^{(c)}(p^2=m_{B_s^*}^2)\Bigg],
\end{eqnarray}
where the constant term reflects the correct renormalization of the 
effective operators. The corresponding $u$-quark loop effect is CKM suppressed 
and therefore neglected.
The $c$-quark polarization operator in Eq.~(\ref{LDBamp}) at $p^2=m_{B_s^*}^2$,
far above the charm-anticharm threshold  can be 
estimated using the dispersion representation 
for the simple $c$-quark loop diagram (``global duality'' approximation)
\beq
 \Pi^{(c)}(p^2)=\frac{p^2}{4\pi^2}\int\limits_{4m_c^2}^{\infty}
\frac{ds}{s(s-p^2-i\epsilon)}\sqrt{1-\frac{4m_c^2}{s}}
\left(1+\frac{2m_c^2}{s}\right)\,.
\label{eq:cloop}
\eeq
Note that the amplitude (\ref{LDBamp}) can be cast in terms of an
effective process-dependent addition $\Delta C_9$ to the coefficient $C_9$ 
in the branching fraction (\ref{SDbrSM_forB}):
\beq
\Delta C_9^{B_s^*\to e^+e^-}=8\pi^2Q_c\left(C_1+\frac{C_2}{3}\right)
\Bigg[-\frac{1}{4\pi^2}\left(\ln\frac{m_c^2}{\mu^2}+
1\right)+ \Pi^{(c)}(m_{B_s^*}^2)\Bigg]\,,
\label{eq:deltaC9}
\eeq
in full analogy with the analysis of the nonlocal charm-loop effect 
in the semileptonic decay, such as $B\to K \ell\ell$ (see, e.g., \cite{KMW}). 
At $m_c=1.3$ GeV, $\mu=4.5$ GeV and at
the Wilson coefficients taken at the same scale: 
$ C_1(\mu)=-0.255 $,  $ C_2(\mu)=1.11 $, 
the numerical calculation yields: 
\beq
\Delta C_9^{B_s^*\to e^+e^-}= 0.11 - 0.47i 
\label{eq:deltaC9}
\eeq
revealing a very small correction to the short-distance coefficient $C_9$.
We therefore skip  the other LD effects, e.g., the one with
the e.m. interaction of the $s$-quark in $B_s^*$.

Finally, to estimate the branching fraction (\ref{SDbrSM_forB})
numerically, we replace $C_9\to C_9+\Delta C_9$,  
use the numerical values for the relevant Wilson coefficients at the scale 
$\mu=4.5$ GeV: 
$C_7^{\rm eff} (\mu) = -0.316$,~
$C_9 (\mu) = 4.293$,
$C_{10} (\mu) = -4.493$,
and employ the QCD sum rule estimate \cite{Gelhausen:2013wia} 
for the $B_s^*$ decay constant $f_{B_s^*} \simeq 250$ MeV. We arrive at the 
 following prediction for the  branching fraction:
\begin{eqnarray}
{\cal B}_{B_s^*\to e^+e^-}= 0.98\times 10^{-11}\,. ~~~
\end{eqnarray}
This estimate implies that already within the SM, one could expect
several events of the type $e^+e^- \to B_{(s)}^{*} \to B_{s}\gamma$ 
to be observed. The signature of the final state is a combination of monochromatic 
low-energy photon and a flavor-violating  $B_s$ meson.

\section{Conclusion.}\label{Conclusions}

We argued that the rare leptonic decays of heavy mesons $H^* = D^*, B^*$ can be probed in the reverse
process of the $e^+e^- \to H^*$ production, provided that the beams of 
$e^+e^-$ collider are tuned in resonance with $m_{H^*}$. 

We calculated relevant transition rates for charm and beauty modes. In the case of charmed mesons 
we paid particular attention to the LD effects, which are 
calculable and found them to exceed the typical SD 
rate by at most one order of magnitude. We also studied several examples of NP scenarios and considered similar production effects for the beauty mesons.
More efforts can be invested in improving the accuracy of 
the estimates presented 
in this paper, by calculating the QCD corrections  to the LD effects 
and by extending the set of NP scenarios sensitive to the processes we 
considered here. 

It would be interesting to note that similar single-charm (single-$b$) final
states can be produced in non-leptonic weak decays of heavy quarkonium
states, such as $J/\psi \to D\pi$. Since heavy charmonium states lie
reasonably far away from the energy region discussed in this paper, these
transitions will not be producing any backgrounds for $e^+e^- \to D^* \to
D\pi$, but can be used to study experimental systematics associated with
such final state. In addition, these transitions are interesting on their
own and will be discussed elsewhere.

Although the experimental setup 
suggested here might look  futuristic, we are convinced that the continuous  progress of 
collider and detector technique will make the  tasks 
suggested in this paper real. We hope that our first 
exploratory estimates will simulate 
dedicated experimental studies of the heavy vector-meson production
in electron-positron collisions. 

{\bf Note added}\\
While we were finishing this paper, the work \cite{Grinstein:2015aua} 
appeared where similar considerations for $B^*$-meson were presented.    

\begin{acknowledgments}
A.A.P. would like to thank Roy Briere for a useful discussion about 
opportunities at BEPCII. A.K. acknowledges valuable comments by
Dirk Seidel.  The work of A.K. and T.M. 
is supported by Deutsche Forschungsgemeinschaft (DFG) Research Unit FOR 1873 
``Quark flavour Physics and Effective Field Theories''.
A.A.P. is supported in part by the U.S. Department of Energy under 
contract DE-SC0007983,  
by Fermilab's Intensity Frontier Fellowship
and also by the Munich Institute for Astro- and 
Particle Physics (MIAPP) of the DFG cluster of excellence "Origin and 
Structure of the Universe." 
A.A.P. is a Comenius Guest Professor at the University of Siegen. 
\end{acknowledgments}



\begin{thebibliography}{99}

\bibitem{CMS:2014xfa} 
  V.~Khachatryan {\it et al.}  [CMS and LHCb Collaborations],
  Nature (2015)
  [arXiv:1411.4413 [hep-ex]].
  
\bibitem{Bobeth:2013uxa} 
  C.~Bobeth, M.~Gorbahn, T.~Hermann, M.~Misiak, E.~Stamou and M.~Steinhauser,
  Phys.\ Rev.\ Lett.\  {\bf 112}, 101801 (2014)
  [arXiv:1311.0903 [hep-ph]].


\bibitem{LHCb:BKll}
  The LHCb Collaboration [LHCb Collaboration],
  LHCb-CONF-2015-002, CERN-LHCb-CONF-2015-002.
  R.~Aaij {\it et al.} [LHCb Collaboration],
  arXiv:1506.08777 [hep-ex].

\bibitem{LHCb:Rk}
  R.~Aaij {\it et al.} [LHCb Collaboration],
  Phys.\ Rev.\ Lett.\  {\bf 113} (2014) 151601
  [arXiv:1406.6482 [hep-ex]].

\bibitem{Artuso:2008vf} 
  M.~Artuso, B.~Meadows and A.~A.~Petrov,
  Ann.\ Rev.\ Nucl.\ Part.\ Sci.\  {\bf 58}, 249 (2008)
  [arXiv:0802.2934 [hep-ph]];
  S.~Fajfer,
  arXiv:1311.6314 [hep-ph];
  G.~Burdman and I.~Shipsey,
  Ann.\ Rev.\ Nucl.\ Part.\ Sci.\  {\bf 53}, 431 (2003)
  [hep-ph/0310076];
S.~Bianco, F.~L.~Fabbri, D.~Benson and I.~Bigi,
  Riv.\ Nuovo Cim.\  {\bf 26N7}, 1 (2003)
  [hep-ex/0309021].

\bibitem{Fajfer:2002gp} 
  S.~Fajfer, P.~Singer and J.~Zupan,
  Eur.\ Phys.\ J.\ C {\bf 27}, 201 (2003)
  [hep-ph/0209250];
  for similar effect in B-decays see
  Y.~G.~Aditya, K.~J.~Healey and A.~A.~Petrov,
  Phys.\ Rev.\ D {\bf 87}, 074028 (2013)
  [arXiv:1212.4166 [hep-ph]].

\bibitem{Zweber:2009qf} 
  P.~Zweber [BESIII Collaboration],
  AIP Conf.\ Proc.\  {\bf 1182}, 406 (2009)
  [arXiv:0908.2157 [hep-ex]].
  
\bibitem{RoyPC}  
 R. Briere, private communications


\bibitem{PDG} 
  K.~A.~Olive {\it et al.}  [Particle Data Group Collaboration],
  Chin.\ Phys.\ C {\bf 38}, 090001 (2014).

\bibitem{Belyaevetal}
  V.~M.~Belyaev, V.~M.~Braun, A.~Khodjamirian and R.~Ruckl,
  Phys.\ Rev.\ D {\bf 51}, 6177 (1995).

\bibitem{Gelhausen:2014} 
  P.~Gelhausen, A.~Khodjamirian, A.~A.~Pivovarov and D.~Rosenthal,
  Eur.\ Phys.\ J.\ C {\bf 74}, no. 8, 2979 (2014)
  [arXiv:1404.5891 [hep-ph]].

\bibitem{Golowich:2009ii} 
  E.~Golowich, J.~Hewett, S.~Pakvasa and A.~A.~Petrov,
  Phys.\ Rev.\ D {\bf 79}, 114030 (2009)
  [arXiv:0903.2830 [hep-ph]].




\bibitem{Burdman:2001tf} 
  G.~Burdman, E.~Golowich, J.~L.~Hewett and S.~Pakvasa,
  Phys.\ Rev.\ D {\bf 66}, 014009 (2002)
  [hep-ph/0112235].

\bibitem{Greub:1996wn} 
  C.~Greub, T.~Hurth, M.~Misiak and D.~Wyler,
  Phys.\ Lett.\ B {\bf 382}, 415 (1996)
  [hep-ph/9603417].


\bibitem{Gelhausen:2013wia} 
  P.~Gelhausen, A.~Khodjamirian, A.~A.~Pivovarov and D.~Rosenthal,
  Phys.\ Rev.\ D {\bf 88}, 014015 (2013)
  [Phys.\ Rev.\ D {\bf 89}, 099901 (2014)]
  [Phys.\ Rev.\ D {\bf 91}, 099901 (2015)]
  [arXiv:1305.5432 [hep-ph]].

\bibitem{Paul:2010pq} 
  A.~Paul, I.~I.~Bigi and S.~Recksiegel,
  Phys.\ Rev.\ D {\bf 82}, 094006 (2010)
  [Phys.\ Rev.\ D {\bf 83}, 019901 (2011)]
  [arXiv:1008.3141 [hep-ph]].
  
\bibitem{SVZ} 
  M.~A.~Shifman, A.~I.~Vainshtein and V.~I.~Zakharov,
  Nucl.\ Phys.\ B {\bf 147} (1979) 385;
448.

\bibitem{BaBarFpi} 
  J.~P.~Lees {\it et al.} [BaBar Collaboration],
  Phys.\ Rev.\ D {\bf 86}, 032013 (2012)
  [arXiv:1205.2228 [hep-ex]].

\bibitem{Petrov:2007gp} 
  A.~A.~Petrov and G.~K.~Yeghiyan,
  Phys.\ Rev.\ D {\bf 77}, 034018 (2008)
  [arXiv:0710.4939 [hep-ph]].

\bibitem{Fajfer:2007dy} 
  S.~Fajfer, N.~Kosnik and S.~Prelovsek,
  Phys.\ Rev.\ D {\bf 76}, 074010 (2007)
  [arXiv:0706.1133 [hep-ph]].

\bibitem{Golowich:2011cx} 
  E.~Golowich, J.~Hewett, S.~Pakvasa, A.~A.~Petrov and G.~K.~Yeghiyan,
  Phys.\ Rev.\ D {\bf 83}, 114017 (2011)
  [arXiv:1102.0009 [hep-ph]].
  


\bibitem{Falk:2001hx} 
  A.~F.~Falk, Y.~Grossman, Z.~Ligeti and A.~A.~Petrov,
  Phys.\ Rev.\ D {\bf 65}, 054034 (2002)
  [hep-ph/0110317].



\bibitem{DstDgamHQ}
  P.~L.~Cho and H.~Georgi,
  Phys.\ Lett.\ B {\bf 296} (1992) 408
   [Phys.\ Lett.\ B {\bf 300} (1993) 410]
  [hep-ph/9209239];
  J.~F.~Amundson, C.~G.~Boyd, E.~E.~Jenkins, M.~E.~Luke, A.~V.~Manohar, J.~L.~Rosner, M.~J.~Savage and M.~B.~Wise,
  Phys.\ Lett.\ B {\bf 296} (1992) 415
  [hep-ph/9209241];
  H.~Y.~Cheng, C.~Y.~Cheung, G.~L.~Lin, Y.~C.~Lin, T.~M.~Yan and H.~L.~Yu,
  Phys.\ Rev.\ D {\bf 47} (1993) 1030\,.
  [hep-ph/9209262];

\bibitem{Colangelo:1993zq}
  P.~Colangelo, F.~De Fazio and G.~Nardulli,
  Phys.\ Lett.\ B {\bf 316} (1993) 555
  [hep-ph/9307330].

\bibitem{DstDgamSR}
  V.~L.~Eletsky and Y.~I.~Kogan,
  Z.\ Phys.\ C {\bf 28} (1985) 155;
  H.~G.~Dosch and S.~Narison,
  Phys.\ Lett.\ B {\bf 368} (1996) 163
  [hep-ph/9510212];
  T.~M.~Aliev, D.~A.~Demir, E.~Iltan and N.~K.~Pak,
  Phys.\ Rev.\ D {\bf 54} (1996) 857
  [hep-ph/9511362];
  J.~Rohrwild,
  JHEP {\bf 0709}, 073 (2007)
  [arXiv:0708.1405 [hep-ph]].

\bibitem{KMW}
  A.~Khodjamirian, T.~Mannel and Y.~M.~Wang,
  JHEP {\bf 1302}, 010 (2013)
  [arXiv:1211.0234 [hep-ph]].

\bibitem{Grinstein:2015aua} 
  B.~Grinstein and J.~M.~Camalich,
  arXiv:1509.05049 [hep-ph].

\end{thebibliography}
\end{document}